\documentstyle[12pt,fleqn]{article}
\input epsf
\def\singlespace {\smallskipamount=3.75pt plus1pt minus1pt
                  \medskipamount=7.5pt plus2pt minus2pt
                  \bigskipamount=15pt plus4pt minus4pt
                  \normalbaselineskip=15pt plus0pt minus0pt
                  \normallineskip=1pt
                  \normallineskiplimit=0pt
                  \jot=3.75pt 
                  {\def\smallskip {\vskip\smallskipamount}}
                  {\def\medskip   {\vskip\medskipamount}}
                  {\def\bigskip   {\vskip\bigskipamount}}
                  {\setbox\strutbox=\hbox{\vrule 
                    height10.5pt depth4.5pt width 0pt}}
                  \parskip 7.5pt
                  \normalbaselines}

\def\doublespace {\smallskipamount=7.5pt plus2pt minus2pt
                  \medskipamount=15pt plus4pt minus4pt
                  \bigskipamount=30pt plus8pt minus8pt
                  \normalbaselineskip=30pt plus0pt minus0pt
                  \normallineskip=2pt
                  \normallineskiplimit=0pt
                  \jot=7.5pt
                  {\def\smallskip {\vskip\smallskipamount}}
                  {\def\medskip   {\vskip\medskipamount}}
                  {\def\bigskip   {\vskip\bigskipamount}}
                  {\setbox\strutbox=\hbox{\vrule 
                    height21.0pt depth9.0pt width 0pt}}
                  \parskip 15.0pt
                  \normalbaselines}

\def\be{\begin{equation}}
\def\ee{\end{equation}}
\def\bea{\begin{eqnarray}}
\def\eea{\end{eqnarray}}

\def\sect #1{\setcounter{equation}{0}}

\begin{document}
\doublespace

\title{\Large{The structure of non-spacelike geodesics in dust collapse}}
\vspace{1.0in}
\vspace{12pt}

\author{S. S. Deshingkar\thanks{shrir@relativity.tifr.res.in}, 
and
P. S. Joshi\thanks{psj@tifr.res.in} \\
\\
Department of Astronomy and Astrophysics\\
Tata Institute of Fundamental Research\\
Homi Bhabha Road, Colaba, Bombay 400005, India. \\}

\maketitle

\newpage


\begin{abstract}
\doublespace

We study here the behaviour of non-spacelike geodesics in 
dust collapse models in order to understand the casual structure 
of the spacetime. The geodesic families coming out, when the 
singularity is naked, corresponding to different initial data
are worked out and analyzed. We also bring out the similarity 
of the limiting behaviour for different types of geodesics 
in the limit of approach to the singularity.

\end{abstract}

\section{Introduction} 

The gravitational collapse of a spherically symmetric dust cloud
is described by the Tolman-Bondi-Lema{\^i}tre (TBL)\cite{TBL} models. These
models have been extensively studied for the validity of the cosmic 
censorship conjecture \cite{dust}. In particular, it is known \cite{INI,GRG} 
now that depending upon the initial conditions, which are defined in terms of
the initial density and velocity profiles from which the collapse
develops, the central shell-focusing singularity at $r=0$ can be 
either a black hole, or a locally or globally naked
singularity. Further, it has been shown recently that this 
singularity is always gravitationally strong \cite{PRD} along a family of
timelike geodesics, indicating 
the physical nature and significance of the singularity. 
There have also been recent studies which examine the stability 
and other aspects in this case \cite{Mena}.

One may thus state that the final fate of gravitational collapse is 
reasonably well-understood for the dust collapse models. The same 
however cannot be said here about the causal structure of the 
singularity, on which several aspects are still not clear, especially
when it is naked. 
It is the behaviour of null geodesics families, closer to the singularity,
that give us an idea of the causal structure there. Most of the earlier 
works on dust collapse have concentrated mainly on establishing the 
occurrence of black holes and naked singularities in TBL models under 
various sets of initial conditions. For the case of a black hole forming 
in dust collapse, the behaviour of null geodesics families around and 
close to the event horizon is rather well-known and fully explored.
However, this is not the case for the behaviour of null geodesics 
in the vicinity of a naked singularity, which is largely
unknown. We take up here a study of the behaviour of radial geodesics 
in the neighbourhood of a naked singularity. We bring out the interesting 
behaviour of null trajectories near naked singularities, and it is 
in this sense that we explore this causal structure here. 
Understanding such issues is clearly going to be important if there 
is any interesting physics to come out due to the existence of a naked 
singularity. Even from the perspective of obtaining a suitable 
formulation of cosmic censorship, such an understanding should be 
helpful and necessary. Our purpose here is to 
investigate the nature and causal structure of this singularity 
by means of examining the behaviour of null as well as 
timelike geodesics families in the TBL models. This provides 
a better understanding of the structure of this singularity, and 
clarifies several related issues.

It is seen from our considerations here that there is 
a single ingoing radial null geodesic (RNG) terminating at the
singularity with a well-defined tangent, and that there is a single RNG 
(the Cauchy horizon) coming out along one direction, and a family of RNGs
coming out in the other direction in all the cases in which the 
singularity is naked. We also show that there is
a family of radial timelike and radial spacelike geodesics coming out of
the singularity with well-defined tangents. We explicitly show these 
results for cases which were not studied so far. In some other cases, 
where one had some idea on the behaviour of the RNGs,
the earlier results can again be reconfirmed from the present 
calculations.

In particular, with our scaling when the physical radius ($R$) 
along the geodesics is proportional to a power of the comoving radius 
($r$) less than 3, 
it is seen that there is only one ingoing radial null geodesic (RNG) 
terminating at the singularity, and one RNG coming out
along the direction of Cauchy horizon as stated above. Further, 
there is a family of radial null and timelike geodesics coming 
out of the singularity along the direction of the apparent horizon. 
In fact, earlier it was claimed in \cite{GRG}, giving an evidence 
for the numerical results obtained there, that in certain cases 
there is a family of RNGs coming out of the 
singularity along the apparent horizon. Here we explicitly 
bring out the existence of such a family, and show that the 
results can be generalized for timelike and spacelike 
geodesics as well.

The plan of the paper is as follows. Section 2 introduces 
the TBL model and the gravitational collapse. In Section 3 we study 
the various singular geodesics, and geodesic families.  
In the concluding section 4 we discuss the overall scenario, 
and the possibility of generalizing these results.

\section{The TBL model and gravitational collapse}

The metric for the TBL models in comoving coordinates is written as,

\be
ds^2= -dt^2 +{{R'^2}\over{1+f}}dr^2 + R^2d\Omega^2.
\ee
The energy-momentum tensor is that of dust,
\be
T^{ij}=\epsilon\delta^{i}_{t} \delta^{j}_{t},\,\,\,
\epsilon=\epsilon(t,r)={F'\over R^2 R'},
\ee
where $\epsilon$ is the energy density, and the area radius
$R =R(t,r)$ is given by
\be
\dot R^2 = f(r) + {F(r)\over R}.
\ee
Here the dot and prime denote partial derivatives with 
respect to the coordinates $t$ and $r$ respectively, and for the 
case of collapse we have  $\dot R <0$. The functions $F$ and $f$ are
called the mass and energy functions respectively, and they are
related to the initial mass and velocity distribution in the cloud.

For a marginally bound cloud ($f=0$), the integration of 
equation (3) gives
\be
t-t_0(r)=-{2R^{3/2}\over 3\sqrt F},
\ee
where $t_0(r)$ is a constant of integration. Using the coordinate freedom 
for scaling the radial coordinate $r$ we set,
\be
R(0,r)=r,
\ee
which gives,
\be
t_0(r)={2r^{3/2}\over 3\sqrt{F}}.
\ee
At the time $t=t_0(r)$ the shell labelled by the coordinate radius $r$ 
becomes singular because the area radius $R$ of the shell becomes 
zero then. Here we consider only the situation 
where there are no shell-crossings in the spacetime. A sufficient 
condition for this is that the density be a 
decreasing function of $r$, which may be considered to be
a physically reasonable  requirement, because for any realistic density 
profile the density should be higher at the center, decreasing 
away from the center. The ranges of coordinates are given by,
\be
0 \le r<\infty,\; -\infty<t<t_0(r).
\ee
The quantity $R'$, which is also needed later 
in the equation of RNGs to check the visibility or otherwise of the 
central singularity, can be written as,
\be
R'={F'R\over 3F} +(1- {rF'\over 3F}) \sqrt{r/R}.
\ee

Though we consider here the marginally bound case $f=0$ for 
the sake of clarity and simplicity, the results are easily 
generalized for non-marginal case.

\section{Non-spacelike geodesics in TBL models}

	In this Section we will first study the basic equations for 
geodesics and write down the root equation  for existence of
geodesics coming out of the singularity. Then we show the existence
of a family of RNGs along the apparent horizon direction, and finally
we study the timelike geodesics in some detail.

\subsection{Basic geodesic equations}

The equations for the radial geodesics in TBL model can be 
written as,
\be
K^t={dt\over dk} = {P\over R},
\label{gs}
\ee
\be
K^r = {dr\over dk}= \pm {\sqrt{P^2 + BR^2}\over RR'} =
\pm {\sqrt{{K^t}^2 + B}\over{R'}}
\ee
where $K^{i}$ is the tangent vector to the geodesic, $k$ is affine 
parameter. In our notation  if there are two signs upper sign always
represents the equation for outgoing geodesic while the lower one represents  
ingoing geodesics. The function $P$ has to satisfy the differential equation,

\be
{dP\over dk}+ (P^2 + BR^2) {\bigg [} {{\dot  R'}\over{RR'}}-{{\dot R}
\over {R^2}} {\bigg ]}
\mp \sqrt{P^2+ BR^2} {{P}\over R^2} - B\sqrt{F/R} =0.
\ee
This gives,
\be
{dt\over dr} =\pm {R'K^t\over \sqrt{{K^t}^2 + B}},
\ee
\be 
{dR\over dr} = R'{\bigg (} 1 \mp \sqrt{{F\over R}} {K^t \over
\sqrt{{K^t}^2 + B}} {\bigg )}.
\label{geoRr}
\ee
In the above equations $B=0$ for null geodesics, $B=-1$ for 
timelike geodesics, and $B=1$ for spacelike
geodesics. We have $dt/dr >0$ for the outgoing geodesics,
and they are represented by the upper sign in above equations.
We introduce new variables $u=r^{\alpha}$ (where $\alpha\ge 1$ is a 
constant), 
$X=R/u$, and write all the quantities 
in terms of these variables. Then, in the limit of approach to the
singularity, using the l'Hospital's rule we can write the
limiting value of $X$ as,
$$
X_{0}\equiv
\lim_{R\to0,u\to0}X
= \lim_{R\to0,u\to0} {{R}\over {u}} 
=\lim_{R\to0,u\to0}
{{dR}\over {du}}= U(X_0,0)
$$
or,
\be
U(X,0) - X \equiv V(X) = 0.
\label{root}
\ee
where $U(X,u)=dR/du$ (along the geodesics).
If the above equation has a real positive root 
$X=X_0$ then the singularity is at least locally naked \cite{INI}.

Near the singularity we assume the form of the mass 
function to be, 
$$
F=F_0r^3 + F_nr^{3+n} + higher \; ordered \; terms.
$$
The first term on the right hand side in the above equation
is required by the condition of regularity of the density 
profile on the initial surface with our scaling.
We need to keep the first two lowest order terms in the mass 
function to get the required behaviour of the geodesics near 
the central singularity. Here we will mainly consider $n<3$, 
i.e. the cases with $\alpha =1+2n/3<3$ in the earlier 
papers \cite{INI}. 
In these cases the singularity is always naked if the density 
decreases as we go away from the centre. The value of the (larger)
real positive root for (\ref{root}) is given by $X_0= (-F_n/2F_0)^{2/3}$ [3],
and it is the same for both ingoing and outgoing geodesics apart from 
the two special cases discussed in \cite{PRD}.

It can be shown (see e.g. Joshi and Dwivedi in Ref. \cite{dust}; see also 
\cite{GRG}) that as $(dV/dX)|_{X=X_0}$ $=h_0<1$ there is 
only one RNG terminating at the singularity with the larger root 
$X_0$ as a tangent. Similarly in these cases it can 
be shown that there is only one ingoing radial null geodesic 
that terminates at the singularity as the $V(X)$ equation is the 
same for both the ingoing and outgoing radial null geodesics, and the 
geodesic equations differ only in higher ordered terms. 
We note that here we mainly study the outgoing geodesics.

\subsection{The family of singular geodesics}

Consider now the collapse with initial densities as 
generated by the mass function as above, with $\alpha<3$. This
corresponds to the initial density profiles as given by
either $\rho(r)= \rho_0 + \rho_1r$, or $\rho(r)= \rho_0 + \rho_2r^2$,
with $\rho_1, \rho_2<0$. We now analyze in these cases the
structure of the geodesics coming out from the singularity
to show that there is such a family of outgoing RNGs 
terminating at the singularity in the past along the  
apparent horizon.

Firstly, let us see in a transparent manner
how such a behaviour is possible for the null geodesics
coming out, and  
terminating in the past at the singularity. For such a 
purpose, take the equation of the curve to be 
$R(r)= b r^{\alpha_1}$ (where $\alpha_1 >1$ and $b$ are constants for
which the value is to be fixed later) to the lowest order, and we see the 
conditions for it to be null
as $r\to 0$.  We assume that $F$ has form $F=F_0 r^3+F_n r^{3+n}$. 
Using the TBL solution we get along this curve,
$$
{dt\over dr} = -{nF_n\over 3{F_0}^{3/2}}r^{n-1} -
(\alpha_1 -1){b^{3/2}\over{\sqrt{F_0}}}r^{{3\over 2}(\alpha_1 -1) -1}
$$
and
$$
R'= br^{\alpha_1 -1} - {nF_n \over 3F_0 \sqrt{b}} r^{n+{1\over2}(1-\alpha_1)}.
$$
Note that we have kept only the lowest order terms which 
will be required for the analysis near the 
central singularity for trajectories which terminate at
the central singularity.
One can take care of more general cases
by (explicitly) keeping $t_0(r)$ in the above equations. 
For the curves to be singular we need  $\alpha_1 >1$ with our scaling.
The conditions for this curve to be null is 
$$
{\Bigg |}{{dt\over dr}\over R'}{\Bigg |} =1.
$$
When $n=3$, we see that this is possible only when $\alpha_1=3$,
and all the powers of $r$ in the numerator and denominator 
become equal and we get the earlier roots equation \cite{TPSJ}, 
i.e. $V(b)=V(X)=0$ for nakedness. 
When $n<3$ we can have two cases. If 
$\alpha_1=3$, then the first term in ${dt/dr}$ and the 
second term in $R'$
dominate, they have equal powers of $r$ and we get $b=F_0$ 
in the limit for the outgoing RNGs. 
This shows that in the limit in the past the outgoing
null geodesics have a similar
behaviour to that of the apparent horizon. The second possible case is
$\alpha_1 =1+ 2n/3$. In this case the two terms in $dt/dr$ have 
an equal power, 
and the two terms in $R'$ also have equal powers. But the power of $r$ 
in $dt/dr$
is less, so we need that the coefficient should vanish and 
again we get the 
earlier roots equation $X_0=(-F_n/2F_0)^{3/2}$ \cite{TPSJ}, and we need higher 
order correction terms to $R(r)$ to 
cancel the powers of $r$. It has been shown earlier that there 
is only one outgoing or ingoing RNG along this direction 
because the value of $h_0 =(dV/dX)|_{X=X_0}$ is less than 1. 
For all other values of $\alpha_1$ we see that the two 
terms in $R'$ and the two terms in $dt/dr$ have different powers,
and also the lowest powers of $r$ in $R'$ and $dt/dr$ are different. 
So we cannot have singular null geodesics with other values 
of $\alpha_1$ possible. Thus we again see that for $n<3$ the 
singularity is always at least locally naked.

Now, we try to check in a simple way when we have
a root to our root equation, whether a family of geodesics can
terminate at the singularity with the given root as a tangent. For
simplicity and clarity we will discuss only the outgoing geodesics,
but with the same method one can easily analyze the ingoing geodesics
as well.

Up to the lowest order, for the singular geodesics, the value of the 
root (tangent) is decided by the self-consistency of 
the differential equations for geodesics; 
the constant of integration (corresponding to the family of geodesics if 
it comes out along the given root direction) can come through only 
higher ordered (additive) terms. To check for such a 
family to exist, we assume that along the geodesics the area radius 
$R$ has the following form (here first we will consider the normal 
root direction),

$$
R \approx X_0r^{\alpha} + Dw(r),
$$
where $D$ is considered to be the constant of integration which can label
different geodesics and $w(r)$ is a function through which the behaviour of
the family comes out.

Now with the form of mass function considered above 
($F=F_0r^3+F_nr^{3+n}$), both the sides of the geodesic equation $(dR/dr)$
can be written to the lowest order of these terms as 
(with $\alpha =1+2n/3$ in this case),
$$
\alpha X_0 r^{\alpha -1} + D{dw(r)\over dr} \approx    
{\bigg [} 1 -\sqrt{{F_0 \over X_0}} r^{1-n/3} + \sqrt{{F_0 \over X_0}}
        {Dw(r)\over 2X_0} r^{-n} {\bigg ]}
$$
\be 
\times{\bigg [} X_0r^{2n/3} +{Dw(r)\over r}
        -{nF_n \over 3F_0\sqrt{X_0}}r^{2n/3} + 
        {nF_n \over 6F_0{X_0}^{3/2}} {Dw(r) \over r} {\bigg ]}.
\ee
So, basically after satisfying the root equation, the differential
equation for $w(r)$ becomes,
\be
\begin{array}{lcr}
 D{dw(r)\over dr} & \approx   &
D {w(r)\over r} {\bigg [} {\bigg (}1-\sqrt{F_0\over X_0}r^{1-n/3}{\bigg )}
{\bigg (}1+{nF_n\over 6F_0X_0^{3/2}}{\bigg )}  \\
 &  & \; ~~~~ \; ~~~~ \; +{1\over 2X_0}\sqrt{F_0\over X_0}
        {\bigg (}X_0 -{nF_n\over 3F_0\sqrt{X_0}}{\bigg )}
        r^{1-n/3}{\bigg ]} \\
 & \equiv & D{dw(r)\over dr}{dU(X,0)\over dX}{\bigg |}_{X=X_0} \\
& = & D{dw(r)\over dr}{\bigg (} 1 + {dV(X)\over dX}{\bigg|}_{X=X_0}{\bigg )}.  
\end{array}
\ee
We define, 
\be
h_0 = 1 + {dV(X)\over dX}{\bigg |}_{X=X_0}.
\ee

Now let us first consider the $n<3$ (i.e. $\alpha <3$) case. 
In this case 
the root is given by, $X_0^{3/2} = -F_n/2F_0$, and so after canceling the 
terms involving the root (as $X_0$ satisfies the root equation for the 
geodesics) we get near the central singularity,
$$
D{dw(r)\over dr} \approx D{w(r)\over r} 
	{\bigg (} 1+{nF_n\over 6F_0X_0^{3/2}}{\bigg )},
$$
i.e.
$$
{dw(r)\over dr} = {w(r)\over r} (1-n/3).
$$
This, after integration gives,
\be
w(r) \propto r^{1-n/3}.
\ee
But as $1-n/3<1$,  $w(r)$ goes to zero slower than $r$, so 
we have only one RNG ($D=0$) coming out along this 
direction.  
Thus we get the similar result to that obtained by 
the earlier method\cite{dust}.

Now let us consider the $n=3$, i.e $\alpha=3$ case. In this case,
when the singularity is naked we have two real positive roots to the
root equation. After cancelling the terms satisfying the root equation 
and retaining the lowest powers of $w(r)$, we get the equation,
$$
D{dw(r)\over dr} = D{w(r)\over r}{\bigg [} {\bigg (}1-\sqrt{F_0\over X_0}
{\bigg )} {\bigg (}1+{nF_n\over 6F_0X_0^{3/2}}{\bigg )}  
+{1\over 2X_0}\sqrt{F_0\over X_0} {\bigg (}X_0 -{nF_n\over 3F_0\sqrt{X_0}}
{\bigg )}{\bigg ]}, 
$$
which after integration gives,
\be
w(r)\propto r^{{\bigg [} {\bigg (}1-\sqrt{F_0\over X_0}
{\bigg )} {\bigg (}1+{nF_n\over 6F_0X_0^{3/2}}{\bigg )}
+{1\over 2X_0}\sqrt{F_0\over X_0} {\bigg (}X_0 -{nF_n\over 3F_0\sqrt{X_0}}
{\bigg )}{\bigg ]}}.
\ee
So, to have a family of geodesics coming out along the tangent 
direction $X_0$ we need,
\be 
h_0\equiv {\bigg [} {\bigg (}1-\sqrt{F_0\over X_0}
{\bigg )} {\bigg (}1+{nF_n\over 6F_0X_0^{3/2}}{\bigg )}
+{1\over 2X_0}\sqrt{F_0\over X_0} {\bigg (}X_0 -{nF_n\over 3F_0\sqrt{X_0}}
{\bigg )}{\bigg ]} >1. 
\ee
This is the same result as shown earlier by Joshi and Dwivedi. 
In this case when the singularity is naked there are two real positive 
roots for the root equation, $V(X)=0$. 
So along one root $h_0 -1$ is negative 
and along the other  it is positive. As the coefficient of
highest power of $X$ in $V(X)$ is negative, 
along the larger root 
$h_0 -1$ is negative while along the smaller root it is positive.
That means along the larger root we have $h_0<1$, and we can have only one 
RNG coming out along that direction (with $D=0$). Along the smaller 
root $h_0>1$ and we have an infinite family of RNGs coming out 
along this direction.

Now let us see what happens in the $n<3$, i.e $\alpha<3$ cases,
along the apparent horizon direction. In this case we need to keep the two
lowest order terms to make sure that we have the apparent horizon as a 
tangent near the singularity. In this case, to the lowest required order,
the equation becomes,
$$
3F_0r^2 + D {dw(r)\over dr} \approx 3F_0r^2 + D{w(r)\over r^3}
{1\over 2F_0}{\bigg (}-{nF_n\over 3F_0^{3/2}}r^{n-1}{\bigg )}.
$$
This, after integration gives,
\be 
w(r) \propto e^{{nF_n \over 6(3-n)F_0^{5/2}}{1\over r^{3-n}}}.
\ee
That means $w(r)$ goes to zero exponentially and we have a family of 
RNGs coming out of the singularity with the apparent horizon kind of 
behaviour.


\begin{figure}[t]
\parbox[b]{6.88cm}
{
\epsfxsize=6.85cm
\epsfbox{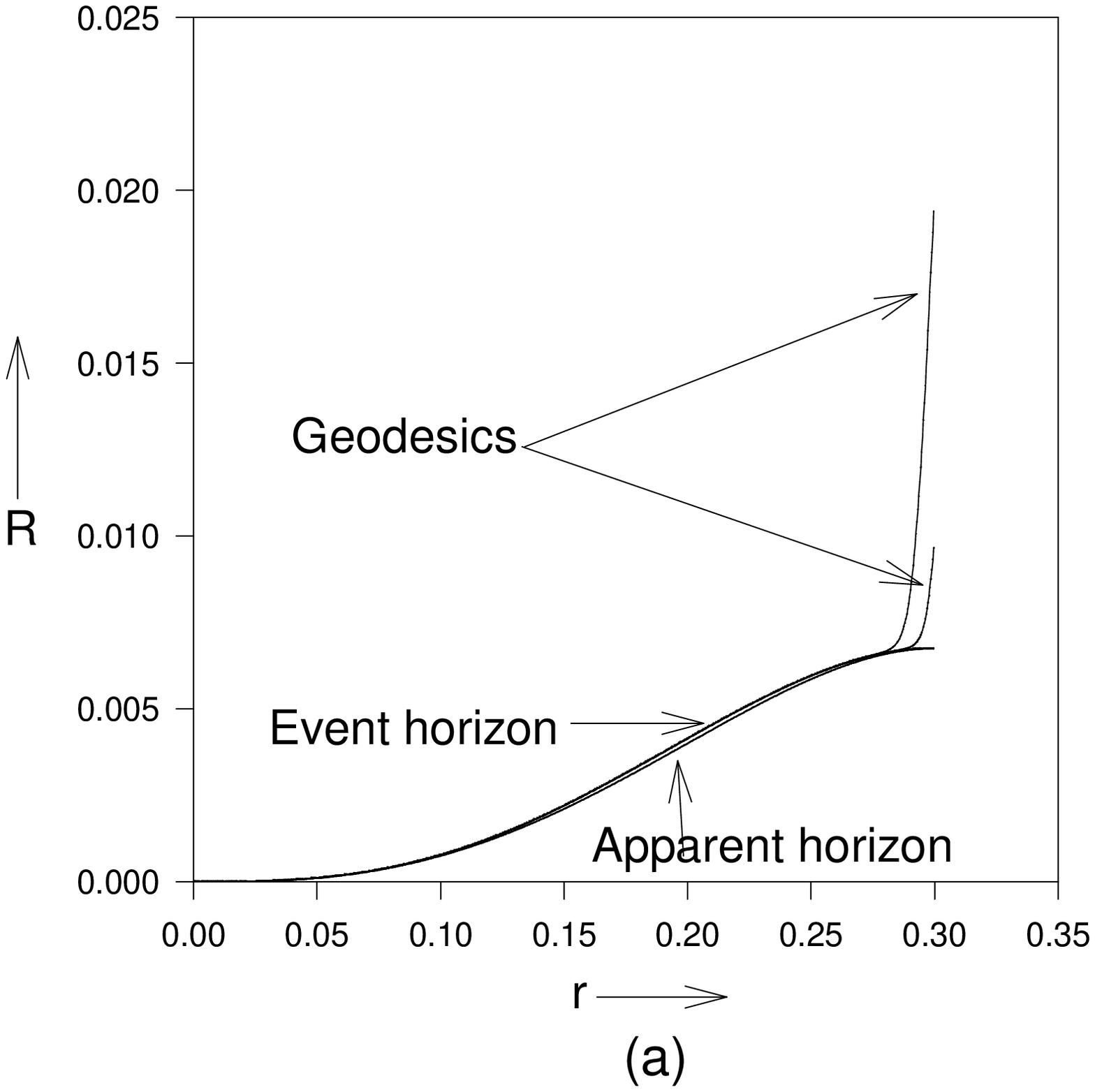}
}
\ \ \
\parbox[b]{6.88cm}
{
\epsfxsize=6.85cm
\epsfbox{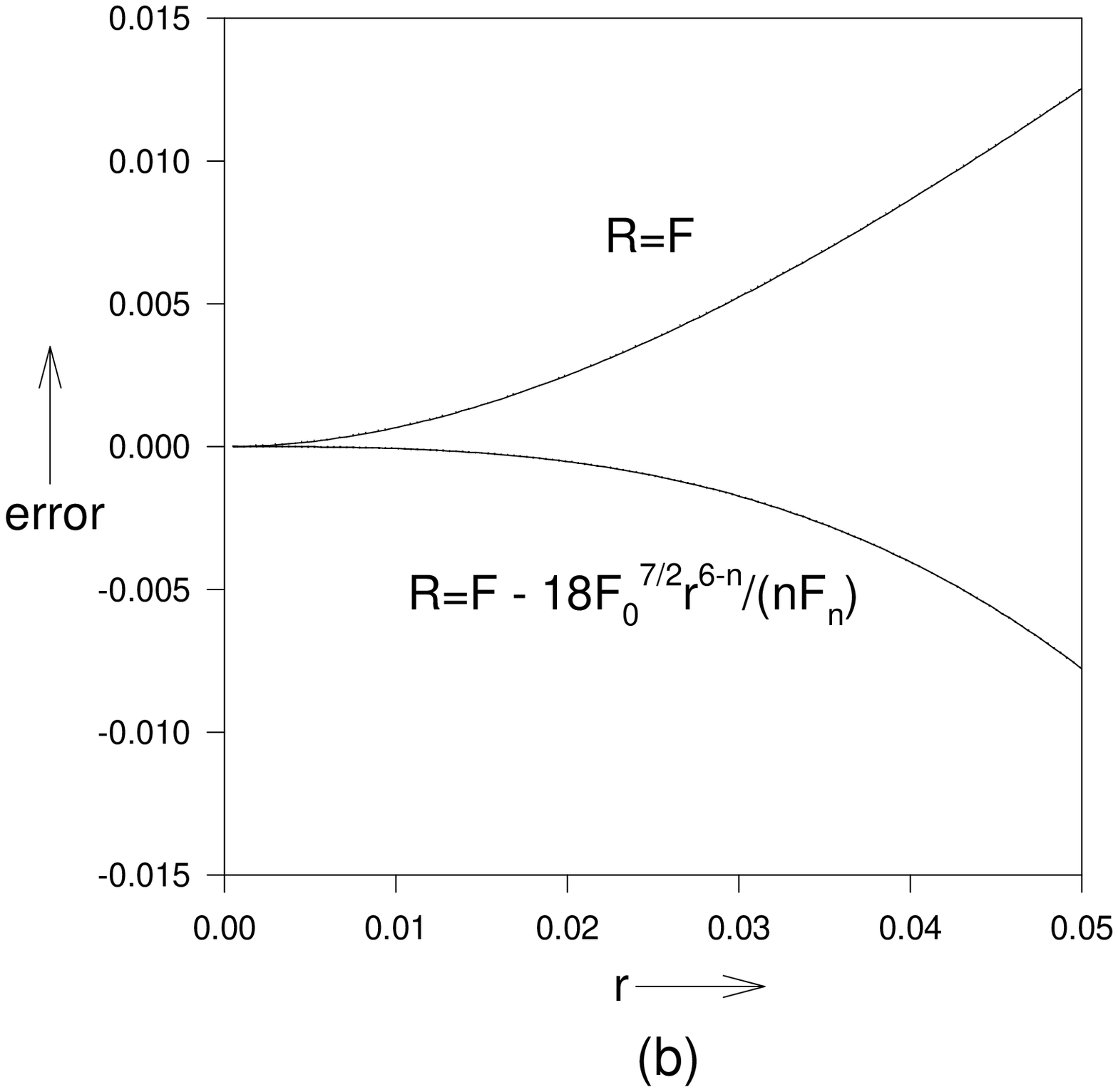}
}
\caption
{ In this models density goes to zero smoothly at the boundary. 
The energy and mass functions are $f=0, \;
F=r^3-2.5r^4$. a) Plots of different outgoing radial null 
geodesics terminating at the singularity in past. We clearly see
the $R=F$ kind of behaviour for these geodesics near the 
singularity. b) graph showing error ${R_{actual} - R_{approximate} 
\over R_{actual}}$ against $r$
near the singularity for $R=F$ curve and the first corrected curve
$R=F -{18 F_0^{3.5}\over nF_n} r^{6-n}$.  }
\end{figure}


\begin{figure}[t]
\ \ \
\parbox[b]{6.88cm}
{
\epsfxsize=6.85cm
\epsfbox{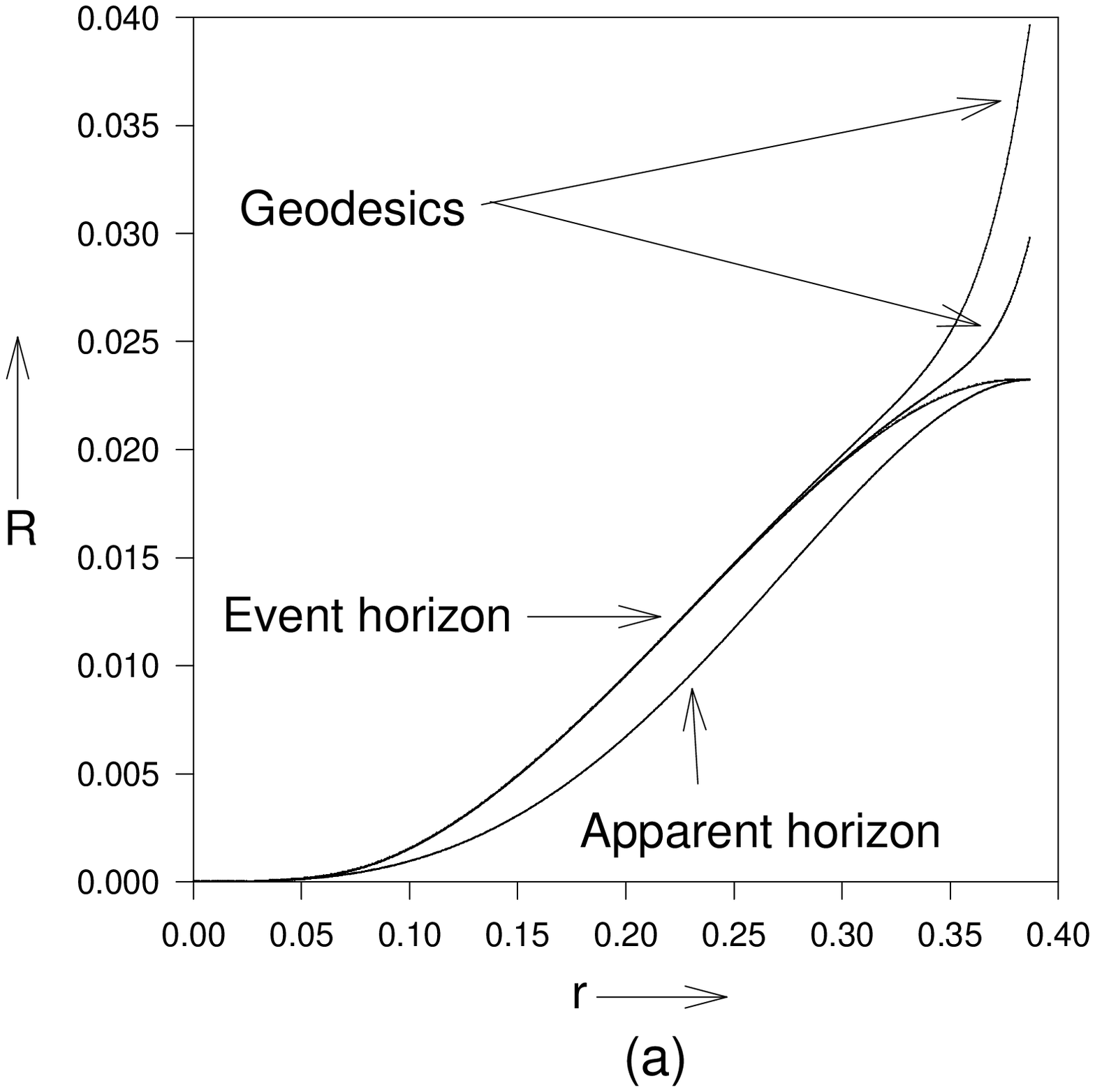}
}
\ \ \
\parbox[b]{6.88cm}
{
\epsfxsize=6.85cm
\epsfbox{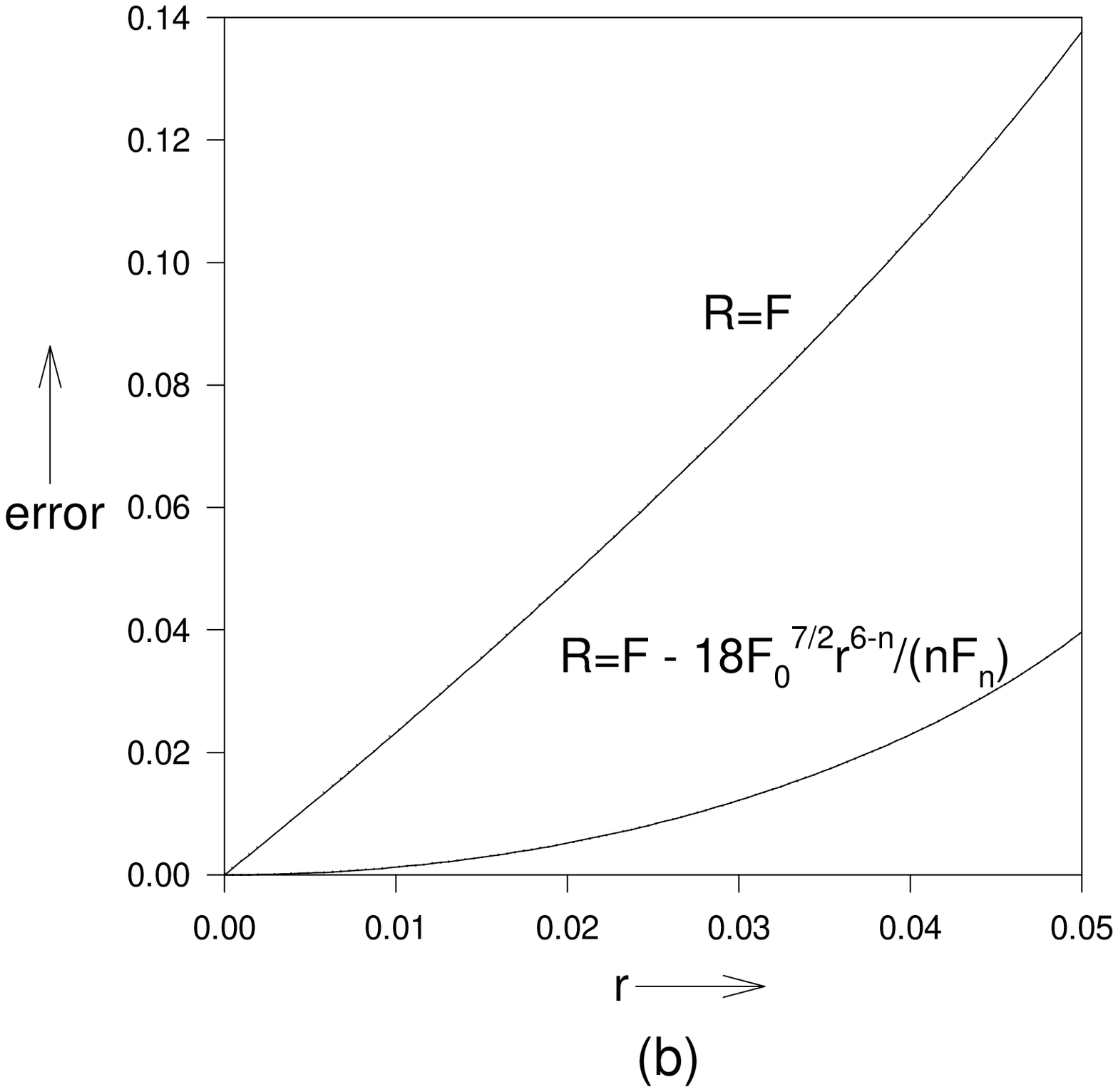}
}

\caption{ Density goes to zero smoothly at the boundary. 
The energy and mass functions are $f=0, \; F= r^3 -2.5 r^5 $. 
a) Plots of different outgoing radial null geodesics terminating
at the singularity. We clearly see
the $R=F$ kind of behaviour for these geodesics near the singularity. 
b) The graph showing error ${R_{actual} - R_{approximate} \over R_{actual}}$ 
against $r$ near the singularity for $R=F$ curve and the first 
corrected curve $R=F-{18 F_0^{3.5}\over nF_n} r^{6-n}$.  }
\end{figure}


The Fig. 1 and Fig. 2 show the plots of the geodesics equations. 
Fig. 1 shows the plots for $F=F_0 r^3 +F_1 r^4$, and Fig. 2 shows the 
plots for
$F=F_0 r^3+F_2 r^5$. We have also plotted the apparent horizon, and the
analytical form of the geodesic equation near the singularity 
with first correction to $F$ [4], i.e. 

$$R= F-{18 F_0^{3.5}\over nF_n} r^{6-n}$$.

It matches very well the actual
geodesic equations near the singularity.
Fig. 1(b) and Fig. 2(b) show the error in analytical 
approximation against
the numerical solution at different values of $r$ near  
the singularity. We see
that the first corrected result to $F$ matches much 
better with the actual 
trajectories near the singularity. The plots also give 
strong numerical
evidence that there is a family of RNGs meeting 
the singularity along the
apparent horizon. The deviation of geodesics
from each other seems to be very drastic as they 
move out. In other words, near the 
center all the singular geodesics converge very fast and
fall on each other as predicted by our analytic calculation.

\subsection{The timelike geodesics}	

Coming to timelike geodesics, by using the set of differential 
equations (\ref{gs}-\ref{root})
for the geodesics, and using self-consistency requirements,
the approximate 
solution for the timelike geodesics near the central singularity
along the normal root direction ($R=X_0r^\alpha$) can be written as,
\be 
K^t={1+C^2\over 2C} \pm r^{1-n/3}{1-C^2\over 2C} 
\sqrt{F_0\over X_0} + higher\; order\; terms,
\ee
\be 
R(t,r)=X_0r^{1+2n/3}, 
\ee
Here  $C\ne1$ is a constant labeling different geodesics 
corresponding to the particles
having different energies at a given point in spacetime, 
and the signs $\pm$ 
represent outgoing and ingoing geodesics.   
The case $C=1$ was discussed in [5].   
So we see that when $C\ne 1$ the timelike and spacelike geodesics have
the same limiting tangent at the singularity as that of the RNGs 
terminating at the singularity. Thus all the geodesics go along 
the same direction,
which we would not have expected to happen typically.

Let us now look for timelike radial geodesics coming
out of the central $(r=0)$ singularity along the apparent horizon,
if they exist,
i.e. with the behaviour
$R=F_0 r^3 +\; higher \; ordered \; terms$. 
Assuming such geodesics 
to exist, and using the set of differential equations 
(\ref{gs} - \ref{root})  
for the timelike
geodesics and solving them in an approximate manner near the singularity,  
we get,
\be
K^t = C\;Exp{\bigg [} {-nF_n\over 3(3-n){F_0}^{5/2}} r^{n-3}{\bigg ]}.
\ee
Here different values of $C$ represents different geodesics 
corresponding to the particles
having different energies at a given point in the spacetime
like in the earlier equation.
We see that $K^t$ and $K^r$ blow up exponentially, 
making $K^t/K^r=1$, and to the 
lowest orders, near
the singularity the $dR/dr$ equation becomes the same as that of the 
null geodesics. Along any other directions
(i.e. apart from the larger root, which is the Cauchy horizon direction,
and the apparent horizon direction, and the 
two special cases discussed earlier\cite{PRD}), we get contradiction 
while solving the set of 
differential equations for the geodesics. 
That means these are the only 
possible behaviours for the geodesics near the singularity. 
And because $K^t$ 
diverges very rapidly near the central singularity, all the arguments 
and calculations given above for null 
geodesics also hold for timelike geodesics, i.e. families
of timelike and null geodesics come out of the 
singularity along the apparent horizon direction near the central  
singularity. 
If one also solves the set of differential equations for 
spacelike geodesics in a self-consistent manner, then
one can see that even the spacelike geodesics have a behaviour 
very similar to that of null geodesics near the central singularity.

Fig. 3 shows the timelike geodesics. 
Within a given range, timelike geodesics starting from the 
singularity can meet the boundary of the cloud at the same time 
with different 
velocities, and there is also a family of such geodesics 
which meet the boundary of the cloud at different times 
as shown in this Figure. 
This implies that in some region, at every point
in the spacetime we have timelike geodesics coming out of the 
singularity with apparent horizon as tangent at the singularity 
and meeting that point with different velocities.

\begin{figure}[t]
\parbox[b]{6.88cm}
{
\epsfxsize=6.85cm
\epsfbox{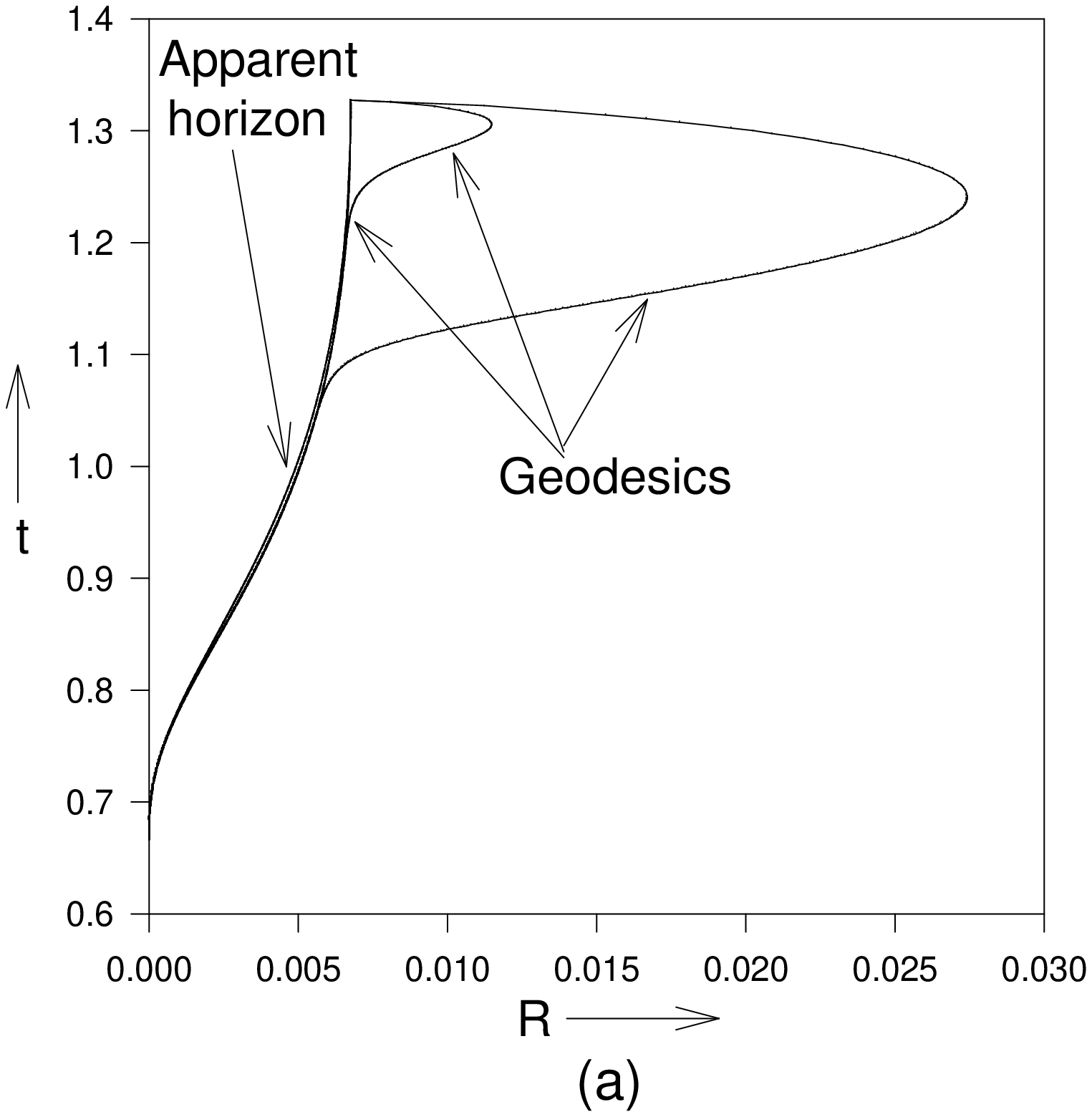}
}
\ \ \    
\parbox[b]{6.88cm}
{
\epsfxsize=6.85cm
\epsfbox{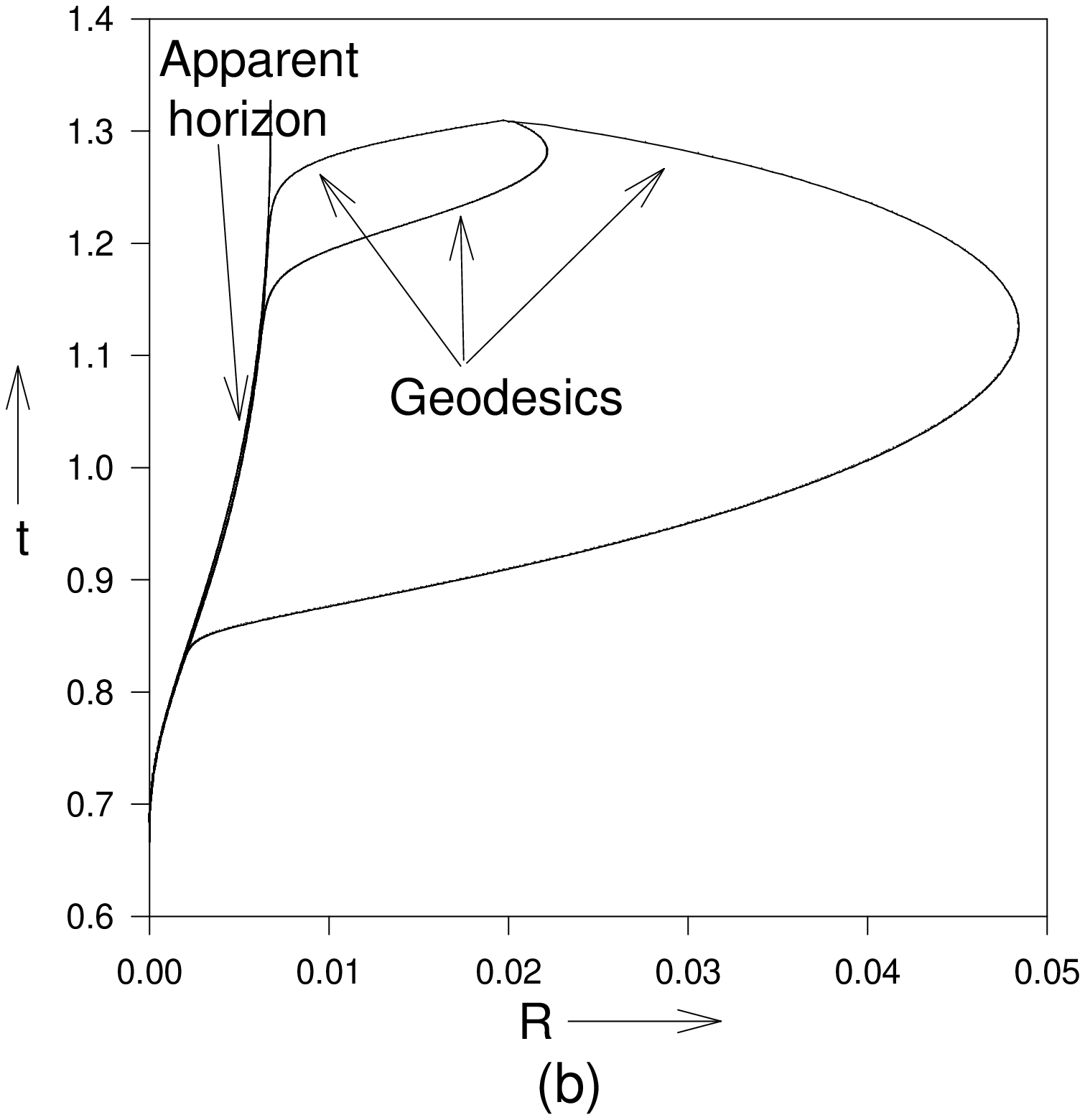}
}

\caption{ Different outgoing timelike geodesics for $f=0 \; F=r^3 - 2.5r^4 $.
Density smoothly goes to zero at the boundary.
All these geodesics show $R=F$ kind of behaviour near the singularity. 
Different timelike geodesics meeting the same point correspond to the
particle having different velocities.
a) Different timelike singular geodesics meeting the apparent horizon at the 
boundary. b) Different timelike singular geodesics meeting R=0.02 at the
boundary.
}
\end{figure}

Though we have considered here the marginally bound case $f=0$ for
the sake of clarity and simplicity, the results are easily
generalized for non-marginal
cases as well. They essentially depend upon the value of the
parameter $\alpha$ which is chosen such that $R'/{r^{\alpha -1}}$ remains
finite in the limit of approach to the singularity along the
trajectories coming out. In the $\alpha <3$ cases, we again get a family of
RNGs coming out of the singularity along the apparent horizon.
This happens because in general 
along the outgoing RNGs (here for simplicity we discuss only null 
geodesics) $dR/dr^3=(1-\sqrt{f+F/R}/\sqrt{1+f})R'/(3 r^2)$, so 
even with $f\ne 0$ we again  have situations with 
$R=F+ M_\delta r^{3+\delta}$, such that the first term in bracket goes 
to zero and the second term blows up but their 
product remains finite, giving the value of the tangent to be $F_0$.

Finally, we make here some remarks in order to clarify the
situation regarding the strength of the naked singularity for the cases
we have considered here. As we indicated earlier, the singularity 
in this model is always gravitationally strong\cite{PRD}. We discuss 
this in some detail below.

The basic idea, in order to examine the strength of a singularity
(naked or otherwise), is to examine the rate of curvature growth
along the non-spacelike geodesics terminating in the singularity, in
the limit of approach to the singularity. There are many criteria
available for this purpose, as we point out below.
According to the definitions of strength of a singularity as
given by Tipler \cite{TIP,CLA1,CLA2} there are {\it two} criteria
given to check the strength.
According to one definition, a divergence condition to be satisfied is,
$R_{ij}V^iV^j$ should go as $1/k^2$ ($k$ is the affine parameter along the 
geodesic, vanishing in the limit of approach to the singularity) as one goes 
to the singularity, along {\it every} geodesic coming out, and also for the 
ingoing ones. 
According to the second definition of Tipler, the singularity 
is strong if this divergence condition is satisfied along at least 
{\it one} trajectory terminating 
in the singularity. We like to use this later definition of strength, 
because there is practically no way available to check the earlier 
definition operationally, because it is nearly impossible to integrate 
all non-spacelike geodesics in most of the collapse scenarios.
Also, considering the complexity of the field equations, for 
most of the collapse models (consider e.g. the well-known case of 
Vaidya-Papapetrou radiation collapse), there is going to be some 
kind of a directional dependence always in the behaviour of the curvature 
growth in the limit of approach to the singularity. A uniform kind of 
curvature dependence in all directions does not seem plausible. Thus, it 
looks extremely reasonable to make the statement that the geodesic ends 
in a strong curvature singularity provided the curvatures grow as per the 
above requirement along it, rather than defining the strength as some 
kind of an absolute property of the singularity.

In fact, there are other criteria as well proposed and 
available to check the strength. For example, according to the criterion 
given by Krolak \cite{KRO} (see e.g. \cite{CLA2}), 
a divergence $R_{ij}V^iV^j$ going 
as $1/k$ is sufficient to call the singularity 
strong. According to this criterion as well, clearly the naked 
singularities considered here 
are all strong curvature singularities along {\it every} 
null geodesic terminating in the singularity. There have also 
been some recent general considerations on strength by Nolan \cite{NOL}, 
according to which practically all shell-focusing singularities 
occurring in spherically symmetric spacetimes are gravitationally 
strong. Taking all these considerations into account, 
it would appear quite reasonable to treat the naked singularities 
considered here to be strong curvature ones, which are physically 
important and not removable from the spacetime in a 
classical manner.

\section{Conclusion}

We have explored here the behaviour of null geodesics in the
vicinity of the naked singularity developing in the gravitational
collapse of a dust cloud. Together with the
earlier information available in this connection (especially in 
the case $n=3$), this provides a rather complete picture of  
how null trajectories traverse away from the naked singularity.

We also see that there are ingoing and outgoing timelike geodesics
terminating at the central naked singularity, and they have a behaviour
very similar to that of the null geodesics when near the 
central singularity.
This is an intriguing phenomenon and because of this in many cases 
it may become difficult to decide whether a given 
vector is null, timelike or spacelike at the central singularity. 
This can happen because normally while doing calculations in such cases 
we finally use some approximations and take limits.

We further showed that there is a family of radial null 
(as well as radial timelike geodesics) 
coming out of the singularity with the same ultimate tangent as that 
of the apparent horizon, when the parameter $\alpha<3$, i.e. the 
geodesics have $R=F$ kind of behaviour as we approach the singularity. 
Very similar results can be obtained for spacelike geodesics also. 
For timelike geodesics in $\alpha=3$ case again we get results quite 
similar to that of null geodesics, as $K^t$ has a power-law divergence 
near the singularity. Even in collapse scenarios more general than dust
we can expect a similar phenomenon, i.e. in the cases corresponding 
to $\alpha <3$, we expect a 
family of geodesics coming out with $R=F$ kind of behaviour as the 
general structure of $dR/dr$ equation looks very similar. 
These results provide us with an insight into the causal structure in 
the vicinity of the naked singularity. 

We thank the referee for helpful comments on the paper.
It is a pleasure to thank I. H. Dwivedi for discussions, and SSD would 
like to thank T. Harada, H. Iguchi, and K. Nakao for many useful
comments.

\end{document}